\documentclass[fleqn, usenatbib]{mnras}

\usepackage[utf8]{inputenc}
\usepackage[T1]{fontenc}
\usepackage{ae,aecompl}

\usepackage{graphicx} 
\usepackage{amsmath}  
\usepackage{amssymb}  

\newcommand{\be}{\begin{equation}}
\newcommand{\ee}{\end{equation}}
\newcommand{\bdm}{\begin{displaymath}}
\newcommand{\edm}{\end{displaymath}}
\newcommand{\bea}{\begin{eqnarray}}
\newcommand{\eea}{\end{eqnarray}}

\newcommand{\msun}{M_\odot}
\def\lsim{\lower.5ex\hbox{$\; \buildrel < \over \sim \;$}}

\usepackage[usenames,dvipsnames]{xcolor}


\begin{document}

\title[A new class of standard sirens] {\textbf Stellar binary black holes in the LISA band: a new class of standard sirens}

\author[W. Del Pozzo et al.]
       {Walter Del Pozzo $^{1,2}$\thanks{E-mail: walter.delpozzo@unipi.it}, Alberto Sesana$^{2}$ \& Antoine Klein$^{3,4,5}$ 
\\
%
$^{1}$ Dipartimento di Fisica ``Enrico Fermi'', Universit\`a di Pisa,  and INFN sezione di Pisa, Pisa I-56127, Italy\\
$^{2}$ School of Physics and Astronomy, University of
Birmingham, Edgbaston, Birmingham B15 2TT, United Kingdom\\
$^{3}$ Department of Physics and Astronomy, The University of 
Mississippi, University, MS 38677, USA\\
$^{4}$ CENTRA, Departamento de F\'isica, Instituto Superior
T\'ecnico, Universidade de Lisboa, Avenida Rovisco Pais 1,
1049 Lisboa, Portugal\\
$^{5}$ CNRS, UMR 7095, Institut d?Astrophysique de Paris, 98 bis Bd Arago, 75014 Paris, France}

\date{\today}

\pagerange{\pageref{firstpage}--\pageref{lastpage}} \pubyear{2016}

\maketitle

\label{firstpage}

\begin{abstract}
  The recent Advanced LIGO detections of coalescing black hole binaries (BHBs) imply a large population of such systems emitting at milli-Hz frequencies, accessible to the Laser Interferometer Space Antenna (LISA). We show that these systems provide a new class of cosmological standard sirens. Direct LISA luminosity distance --$D_l$-- measurements, combined with the inhomogeneous redshift --$z$-- distribution of possible host galaxies provide an effective way to populate the $D_l-z$ diagram at $z<0.1$, thus allowing a precise local measurement of the Hubble expansion rate. To be effective, the method requires a sufficiently precise LISA distance determination and sky localization of a sizeable number of BHBs, which is best achieved for a 6-link detector configuration. We find that, for a BHB population consistent with current fiducial LIGO rates, the Hubble constant $H_0$ can be determined at the $\sim$5\% and $\sim$2\% level (68\% confidence) assuming two and five million Km arm-length respectively.
\end{abstract}


\section{Introduction}
The first direct gravitational wave observations by Advanced LIGO \citep{2016PhRvL.116f1102A,2016PhRvL.116x1103A} opened a new era in observational astronomy. The two detected events (GW150914 and GW151226) plus a third candidate (LVT151012) are all consistent with merging black hole binaries (BHBs) \citep{2016arXiv160604856T}, and provide a formidable laboratory for astrophysics \citep{2016ApJ...818L..22A} and fundamental physics \citep{2016PhRvL.116v1101A}. As first pointed out by \cite{2016PhRvL.116w1102S}, the high merger rates inferred by the LIGO detections imply an unexpectedly vast population of fairly massive BHBs emitting at milli-Hz frequencies, accessible to space based interferometers. This is an appealing prospect for the Laser Interferometer Space Antenna \citep[LISA][]{2013arXiv1305.5720C} planned by the Europeans Space Agency (ESA), with a possible launch window within 2029-2035.

Besides observing several BHBs few years before they merge in the LIGO band, thus realizing the promise of multi-band gravitational wave (GW) astronomy \citep{2016PhRvL.116w1102S,2016PhRvL.116x1104B,2016PhRvL.117e1102V}, many more systems will be observed as slowly inspiralling sources around $5-10$mHz, hundreds of years before their final coalescence \citep[see also][]{2016MNRAS.462.2177K}. This new population might prove useful in a number of astrophysical and cosmological contexts. For example, eccentricity measurements which are possible in the LISA band \citep{2016PhRvD..94f4020N} can be used to discriminate among different astrophysical formation channels \citep{2016arXiv160609295N,2016arXiv160609558B}. In this paper we investigate the prospects of using LISA measurements of BHBs to constrain cosmological parameters, and in particular to provide an independent local measurement of the Hubble expansion rate $H_0$. This is of paramount importance in light of the recent tension between local and cosmological determination of this parameter. If we write $h=H_0/(100$km s$^{-1}$ Mpc$^{-1})$, a recent analysis of type Ia supernovae with improved distance calibrators yields $h=0.732\pm0.017$ (1$\sigma$ error) \citep{2016ApJ...826...56R}, largely inconsistent with Planck measurements of $h=0.678\pm 0.009$ \citep{2015arXiv150201589P}. Although local measurements might suffer from cosmic variance effects, \cite{2016ApJ...826...56R} showed that this is unlikely to bias the measurement of $h$ at more than 1\%. Therefore, a completely independent local measurement, not affected by the same distance calibration systematics, might prove useful in resolving the tension.

The argument of using GW sources as 'standard sirens' for cosmology measurements dates back thirty years \citep{1986Natur.323..310S} and goes as follows. Contrary to electromagnetic (EM) probes, GW observations provide a direct measurement of the luminosity distance --$D_l$-- to the source. If the source redshift can then be determined through the detection of an associated EM counterpart, one can populate the $D_l-z$ diagram, thus extracting the cosmological parameters that provide the best fit to the observed points. This idea has been extensively exploited in different contexts both for ground based \citep[e.g.][]{2012PhRvD..86d3011D,NissankeEtAl:2010} and space based detectors \citep{2005ApJ...629...15H,2016JCAP...04..002T}. It has been noted, however, that for specific classes of sources distinctive EM counterparts might be difficult to identify, like in the case of massive BHBs (MBHBs) at cosmological distances, or might not be expected in first place, like in the case of extreme mass ratio inspirals (EMRIs), i.e. compact objects inspiralling into a massive BH. Even in this case, however, one can still take advantage of the clustering properties of galaxies, an idea that was first proposed by \cite{2008PhRvD..77d3512M} in the context of EMRIs observations and then applied by \cite{2011ApJ...732...82P} to MBHBs. To illustrate the basic idea, let assume that LISA provides perfect measurements of $D_{l,i}$ on a set of GW sources $i=1,...,n$. If cosmological parameters are unknown within a given prior range, each measured $D_{l,i}$ will translate in a possible $\Delta{z_i}$ range for the host galaxy. Therefore, each event will have $j=1,....,m_i$ possible hosts at different redshifts within $\Delta{z_i}$. One can then cross-correlate the $N$ events and search for the unique set of cosmological parameters that aligns one galaxy in each $\Delta{z_i}$ on a single $D_l-z$ relation. LISA measurement errors and the presence of peculiar velocities add significant complications to the method, adding uncertainties in the determination of the parameters, but the basic concept remains unchanged.

Here we apply the method of \cite{2008PhRvD..77d3512M} and \cite{2012PhRvD..86d3011D} to the putative population of stellar BHBs observable by LISA. The idea of using these systems as standard sirens has independently been proposed by \cite{2016arXiv160907142K}. Instead of using a statistical method, they estimated the number of BHBs that can be localized well enough so that only one galaxy falls in the GW measurement error cube. Based on analytic scaling relations for the average sky location and distance accuracy, they concluded that LISA might measure $H_0$ to a few\% precision. This result, however, relies on average estimates of quantities that vary by orders of magnitude from one source to another. Here we show, in fact, that association of the BHB with a single galaxy host is generally possible only in a handful of cases, and statistical methods to reconstruct the $D_l-z$ relation are of crucial importance. 

The paper is organized as follows. In Section \label{sec:theo} we present the analytical framework and introduce the statistical tools needed for the analysis. In Section \ref{sec:astromodel} we construct the BHB population observed by LISA and the associated distribution of galaxy counterparts, which are the necessary input to our analysis. Results on the determination of $h$ are presented in Section \ref{sec:results} and final remarks are left to Section \ref{sec:conclusions}.

\section{Theoretical framework}
\label{sec:theo}

In this section, we summarise the general mathematical treatment given in \cite{2012PhRvD..86d3011D,2015arXiv150606590D}
and specialise it to the case in which the redshifts of the potential galaxy hosts are given by the sum of the cosmological redshift and 
the peculiar one given by the galaxy proper motion.

Consider a set of $n$ GW observations $\vec{d}\equiv d_1,\ldots,d_n$ and an all sky galaxy catalogue $\vec{g}$. 
Each galaxy $g_j$ in the galaxy catalogue is defined (for simplicity) in terms of its right ascension and declination coordinates 
on the celestial sphere $\varphi_j,\theta_j$ and its observed redshift $z_j$ which is given by the sum of the cosmological redshift 
and the apparent redshift due to the galaxy peculiar velocity. Under the assumption that each GW event is located in one 
of the galaxies in the Universe, we want to construct the posterior probability distributions for an arbitrary set of cosmological 
parameters $H\equiv H_0, \Omega_m, \ldots$. 
From Bayes' theorem: 
\begin{align}
p(H|\vec{d},I) \propto p(H|I)p(\vec{d}|H,I) 
\end{align}
where $p(H|I)$ is the prior probability distribution for the cosmological parameters $H$ and  $p(\vec{d}|H,I)$ is the (quasi-)likelihood function 
for the observations $\vec{d}$. We indicate by $I$ all background information that is relevant for the problem at hand. 
Since each GW observation is independent, the likelihood factorises in the product of the likelihoods for the individual events:
\begin{align}\label{eq:joint-like}
  p(\vec{d}|H,I) = \prod_{i=1}^n p(d_i|H,I)\,.
\end{align}
The (quasi-)likelihood for a GW event is obtained by marginalisation over its celestial coordinates $\theta,\varphi$, luminosity distance $D_L$ and redshift $z$. Define $\mathbf{ x}\equiv \{D_L,z,\theta,\varphi\}$:
\begin{align}\label{eq:marg-like1}
p(d_i|H,I) = \int d \mathbf{x}\, p(d_i|\mathbf{x},H,I)p(\mathbf{x}|H,I)\,.
\end{align}
We are going to assume that the integral over the celestial coordinates can be factorised and performed analytically\footnote{The marginalisation over sky coordinates corresponds to the selection of 
only the galaxies within the cone subtended by the measured solid angle}. We are left then with: 
\begin{align}\label{eq:marg-like}
p(d_i|H,I) = \int d D_L dz\, p(d_i|D_L,z,H,I)p(D_L|z,H,I)p(z|H,I)\,.
\end{align}
As in \citet{2012PhRvD..86d3011D}, 
\begin{align}
p(D_L|z,H,I) = \delta(D_L-D_L(z,H))
\end{align}
with $D_L(z,H)$ given by \citet{hogg:1999}. Under the simplifying assumption that the likelihood 
for the GW event is given by a Gaussian distribution in luminosity distance $D_L$ centred around some mean value 
$\bar{D}_L$ and with standard deviation $\sigma_D$, after integrating over $D_L$, we get 
\begin{align}
p(d_i|D_L(z,H),z,H,I) \propto \exp{\left[-\frac{1}{2}\left(\frac{D_L(z,H) - \bar{D}_L}{\sigma_D}\right)^2\right]}\,.
\end{align}
Finally, we need to define the prior over the GW event redshift $p(z|H,I)$. 
Differently from \citet{2012PhRvD..86d3011D}, we do not simply choose a sum of delta functions over the individual 
galaxy redshifts, but we account also for the contribution to the redshift due to the proper motion. We choose a prior of the form
\begin{align}
p(z|H,I)\propto \sum_{j=1}^K \exp{\left[-\frac{1}{2}\left(\frac{z_j - z}{\sigma_{pv}}\right)^2\right]}\,
\end{align}
with $\sigma_{pv} = 0.015$. This value is chosen by inspection of the distribution of line of sight velocities in the Millennium simulation.
We finally obtain the likelihood for a single GW event and the corresponding potential galactic hosts:

\begin{align}\label{eq:like-single}
p(d_i|H,I) &= \int_{z_{min}}^{z_{max}}  dz \exp{\left[-\frac{1}{2}\left(\frac{D_L(z,H) - \bar{D}_L}{\sigma_D}\right)^2\right]}\times\nonumber \\
&\sum_{j=1}^K \exp{\left[-\frac{1}{2}\left(\frac{z_j - z}{\sigma_{pv}}\right)^2\right]}\,.
\end{align}
The limits of integration $z_{min},z_{max}$ are chosen as described in Sec.~\ref{sec:counteparts} for the host galaxies selection. 

\subsection{Numerical implementation}

We explore Eq.~(\ref{eq:joint-like}) and Eq.~(\ref{eq:like-single}) numerically using a parallel implementation of the Nested sampling algorithm\footnote{https://github.com/johnveitch/cpnest}
specifically prepared by the authors and adapted from the algorithm used in \cite{dpw:2016}. 
For each of the scenarios presented below, we compute posterior probability distributions for the reduced Hubble constant $h = H_0/100\cdot km\,s^{-1}\,Mpc^{-1}$ and for the matter density $\Omega_m$.
In addition to these parameters, because of the peculiar velocities of the galaxies in the Millennium simulation, 
we also compute posterior probability functions for the each GW event includes in the analysis. 
We choose uniform prior distributions in $[0.1,1.5]$ for $h$ and $[0.04,2.0]$ for $\Omega_m$. The prior for each GW event redshift is chosen to be uniform between $[z_{min},z_{max}]$. 
We explore the parameter space using an ensemble of 1000 Live points. At every iteration of the Nested Sampling algorithm, 
we generate a new Live point evolving the worst likelihood one via a Markov Chain Monte Carlo (MCMC) \citep[see][]{Veitch:2014wba}. 
We limit the maximum number of MCMC step to 5000.
The posterior distribution for $h$ and $\Omega_m$ is constructed from the the individual events marginal likelihoods in the same way described in \citet{2015arXiv150606590D}: 
once samples from the marginal likelihood are available, we use a Dirichlet Process Gaussian Mixture model (DPGMM) \citep{Ferguson1973a,Hjort2010a} to obtain a density estimate 
of the likelihood density. The joint distribution is then computed by straightforward multiplication.

\section{Astrophysical model}
\label{sec:astromodel}

The practical application of the formalism developed in the previous section requires the numerical evaluation of the likelihood given by equation (\ref{eq:like-single}). This requires the specification of several astrophysical inputs:
\begin{enumerate}
\item a population of GW sources defining the vector of observed events $d_1, ..., d_n$;
\item an estimate of the errors in the determination of the source sky location and luminosity distance, which define the redshift support of possible galaxy counterparts in Eq.~(\ref{eq:like-single});
\item the ensemble of galaxies $g_{i,1},...g_{i,m}$ populating the error box of each GW event $d_i$, and their apparent redshifts. 
\end{enumerate}
In the following, we describe how we construct each relevant quantity.

\subsection{Black hole binary population}

To generate a cosmic population of inspiralling BHBs in the milli-Hz window, we follow the procedure described in \cite{2016PhRvL.116w1102S}. For the given BHB mass function, one can compute the associated chirp mass probability distribution $p({\cal M}_r)$, where ${\cal M}_r=(M_{1,r}M_{2,r})^{3/5}/(M_{1,r}+M_{2,r})^{1/5}${\footnote{In the following, the subscript $r$ refers to quantities computed in the BHB rest frame. Those are related to quantities measured on Earth as $M_r=M/(1+z)$, $f_r=f(1+z)$, $t_r=t/(1+z)$.}}. The merger rate density per unit mass can be then written as
\begin{equation}
  \frac{d^2n}{d{\cal M}_rdt_r}={\cal R}\times p({\cal M}_r),
  \label{eqrate}
\end{equation}
where ${\cal R}$ is the total rate units of yr$^{-1}$Gpc$^{-3}$. The number of BHBs per unit mass, redshift and frequency is therefore given by:
\begin{equation}
  \frac{d^3N}{d{\cal M}_rdzdf_r}=\frac{d^2n}{d{\cal M}_rdt_r}\frac{dV}{dz}\frac{dt_r}{df_r},
  \label{eq1}
\end{equation}
where $dt_r/df_r$ is the binary frequency evolution implied by the standard quadrupole emission formula (we assume circular binaries), and the volume shell $dV/dz$ is given by a fiducial $\Lambda$CDM cosmology with $h=0.73,\,\Omega_m=0.25,\,\Omega_\Lambda=0.75$. Although those parameters are not up to date, this is irrelevant for the purpose of our exercise. The choice has been made to be consistent with the simulated universe employed to select galaxy counterparts, as described in Section \ref{sec:counteparts} below.

A Monte Carlo sampling of the differential distribution given by equation (\ref{eq1}) in the relevant frequency ($10^4{\rm Hz}<f<0.1{\rm Hz}$) and redshift ($z<0.5$) range provides the BHB population observable in the LISA band. All we need to specify is the intrinsic rate ${\cal R}$ and the BHB mass function, defining $p({\cal M}_r)$. LIGO observations imply a cosmic BHB merger rate in the range ${\cal R}=9-240$ yr$^{-1}$Gpc$^{-3}$ \citep[90\% confidence][]{2016arXiv160604856T}. With different assumptions in the underlying BHB mass function leading to different rate estimates. For the results shown in Section \ref{sec:results}, we considered a log-flat rest-frame BHB mass function in the range $5\msun<M_{1,2,r}<100\msun$ and three values of the intrinsic merger rates ${\cal R} \approx 12-34-70 $yr$^{-1}$Gpc$^{-3}$. Those roughly correspond to the $10-50-90$-th percentile of the rate probability distribution for this mass model \citep[see figure 10 in][]{2016arXiv160604856T}, thus providing a conservative, fiducial and optimistic scenario. We stress that, as shown in \cite{2016PhRvL.116w1102S}, the number of BHBs in the LISA band implied by the LIGO results is almost independent on the details of the BHB mass function. For example, a Salpeter mass function, although resulting in an higher estimate of ${\cal R}$, would essentially give the same number of detections in the LISA band. Therefore, our results do not strongly depend on the specific assumptions of a log-flat mass function.

\subsection{LISA observations}
We consider two LISA detector configurations, namely N2A2M5L6 and N2A5M5L6 in the nomenclature of \cite{2016PhRvD..93b4003K}. Those correspond to detector baselines with low frequency noise set to the original LISA requirement (N2), two and five million kilometre arm-length (A2 and A5, respectively), a nominal five year mission lifetime (M5), and six laser links (L6). The first assumption is based on the reported performance of the LISA Pathfinder spacecraft \citep{2016PhRvL.116w1101A}, whereas the remaining options are set following the recommendation of the Gravitational Observatory Advisory Team to the European Space Agency {\footnote{http://www.cosmos.esa.int/web/goat}}. Since the two configurations only differ by their arm-length, in the following we will refer to them simply as A2 and A5.

Individual signal-to-noise ratios (S/N) and estimates of the relevant parameters are produced using the Fisher information matrix (FIM) approximation, and the analytical sensitivity curves provided by \cite{2016PhRvD..93b4003K}.
The GW signals are modelled using 3.5 Post Newtonian precessing spinning waveforms for circular adiabatic inspirals based on the shifting asymptotic method (SUA) described in \citep{2014PhRvD..90l4029K}. For a given detector configuration and BHB population model we therefore generate a catalogue of $i=1,...,N$ sources observed with S/N$>8$. Each observed system is defined by a vector of parameters $\vec{\lambda}_i$ (including masses, spins, sky location, luminosity distance, etc.) with associated errors $\Delta\vec{\lambda}_i$ estimated in the FIM approximation. 

For the purpose of host identification, errors on distance $\Delta{D_l}$ and sky localization $\Delta{\Omega}$ are of paramount importance. The latter is defined as $\Delta \Omega=2\pi \sin\theta \sqrt{\Delta \theta \Delta \phi - (\Sigma^{\theta\phi})^2}$, where $\theta$ and $\phi$ are the latitude and longitude of the source in ecliptic coordinates and $\Sigma^{\theta\phi}$ in the correlation between those two quantities. For the reconstruction of the $D_l-z$ diagram, only events with a decent $D_l$ estimate are useful. We therefore select systems with $\Delta{D_l}/D_l<0.2$. Furthermore, since we are primarily interested in an independent measurement of $h$, we restrict our analysis to events at a distance corresponding to $z<0.1$ in the chosen cosmology. We checked, in any case, that only few events at $z>0.1$ have $\Delta{D_l}/D_l<0.2$ and should not significantly improve on the measurement of $h$.

\begin{figure}
\begin{center}
\includegraphics[width=8.5cm]{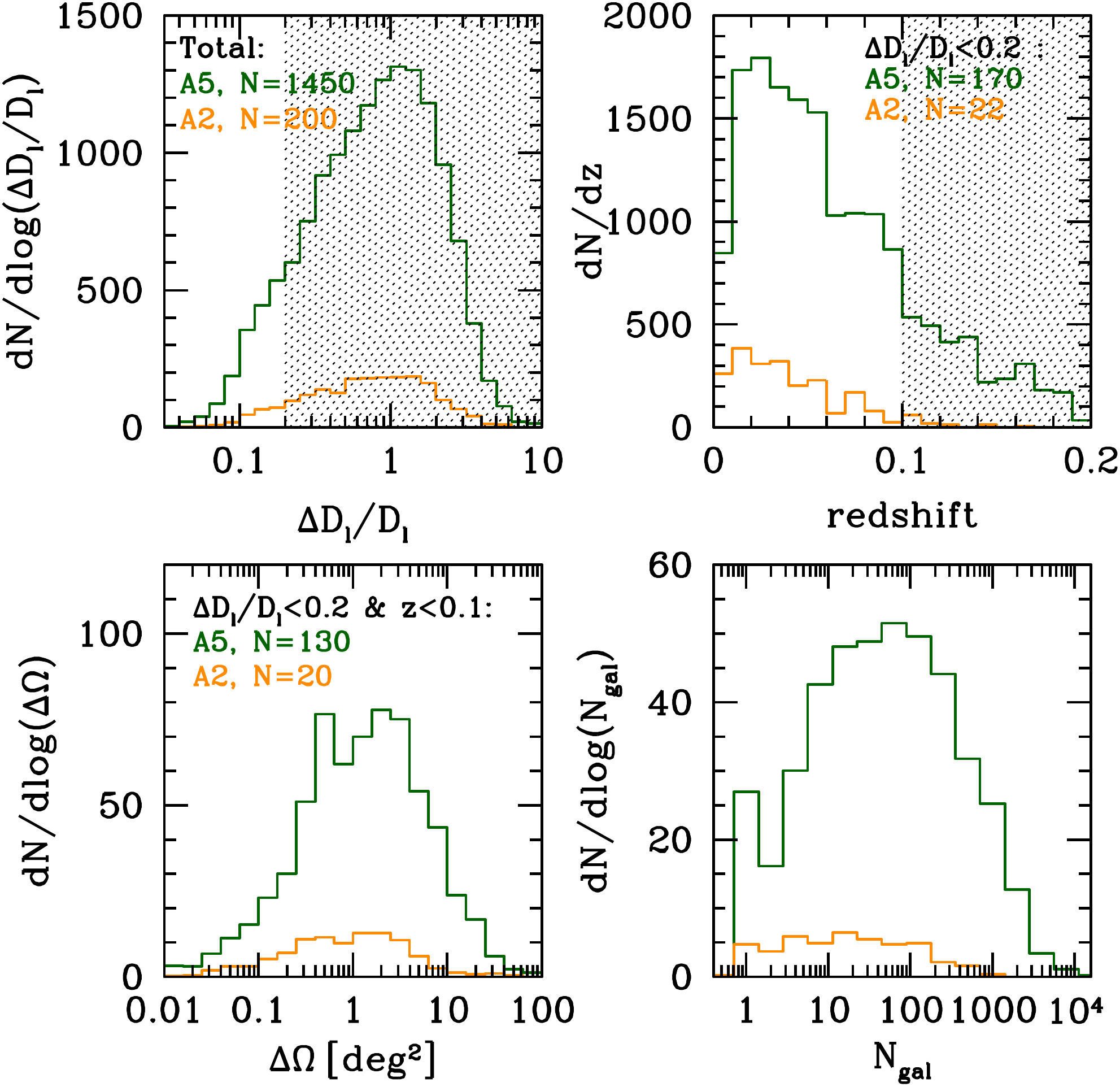}
\caption{Properties of the average BHB population observed by LISA assuming the A5 (green) and the A2 (orange) baselines. Top left: distribution of luminosity distance errors for all sources with S/N$>8$. Top right: redshift distribution of the subset of sources with $\Delta{D_l}/D_l<0.2$. Bottom left: sky localization error distribution on the subset of sources at $z<0.1$. Bottom right: distribution of the number of possible galaxy hosts in the error-box defined by LISA measurement errors and the assumed prior range in the cosmological parameters. Numbers always refer to the integral of the distributions plotted in each panel.}
\label{fig:pe}
\end{center}
\end{figure} 

The cuts in $\Delta{D_l}/D_l$ and $z$ are shown in Fig. \ref{fig:pe}, where we show parameter estimation results averaged over 20 Monte Carlo realizations of the BHB population with ${\cal R}$ sampled from the posterior distribution derived by the LIGO observations. The top right panel shows the $\Delta{D_l}/D_l$ distribution. Because of the relatively low S/N and of the inclination-distance degeneracy, distance is essentially undetermined for about 50\% of the sources, and only about 11\% of them ($170/1450$ and $22/200$ in the A5 and A2 case respectively) satisfy the requirement $\Delta{D_l}/D_l<0.2$. Most of those sources are at $z<0.1$ (top left panel), and they can be localized with an accuracy of about 1deg$^2$ (bottom right panel). The average number of sources that can be exploited for cosmological inference is therefore about 130 for the A5 configuration and only 20 for the A2 one.   

\subsection{Host galaxy selection}
\label{sec:counteparts}
For a given cosmology, a LISA measurement $D_l\pm \Delta{D_l}$ translates into a specific $z\pm \Delta{z}$. This conversion is complicated by two facts. Firstly, since our goal is to determine the cosmological parameters, we need to increase our $\Delta{z}$ estimate to account for the prior range in those parameters. In practice, the minimum and maximum redshifts, $z^-,z^+$, compatible with a given distance measurement and some cosmological prior are implicitly given by:  
\begin{equation}
  D_l-\Delta{D_l}=(1+z^-)D_0\int_0^{z^{-}} \frac{dz'}{F^-(z')},
\end{equation}  
\begin{equation}
  D_l+\Delta{D_l}=(1+z^+)D_0\int_0^{z^{+}} \frac{dz'}{F^+(z')},
\end{equation}  
where $D_0=9.26\times 10^{27}$ cm and $F(z)=hE(z)$, with $E(z)$ defined in \citet{hogg:1999}. $F^-(z)$ and $F^+(z)$ are the specific realizations of the function $F(z)$ that minimize and maximize the $D_l-z$ conversion within the assumed cosmological prior for the parameters $h, \Omega_m, \Omega_\Lambda$. As specified above, in this investigation we assume very conservative prior $h\in[0.6,0.85]$ and $\Omega_m\in[0.04,1]$ (fixing $\Omega_m+\Omega_\Lambda=1$). Therefore $F^-(z)=0.6[0.04(1+z)^3+0.96]^{1/2}$ and $F^+(z)=0.85(1+z)^{3/2}$. Secondly, because of peculiar velocities $v_p$, the apparent redshift $z_{\rm obs}$ of the GW host might be different from the cosmological redshift $z$ determined by the Hubble flow, thus adding a further uncertainty term $\Delta{z}_{v_p}=(1+z)v_p/c$ to the estimated redshift. To account for this effect, for each event defined by $D_l\pm \Delta{D_l}$, we assign a possible host redshift range $[z^{-}-\Delta{z^-_{v_p}}\, ,\, z^{+}+\Delta{z^+_{v_p}}]$. We assume $v_p=500$ km s$^{-1}$, consistent with the standard deviation of the radial peculiar velocity distribution observed in the Millennium run \citep{2005Natur.435..629S}.

The support $\Theta$ that enters in the evaluation of the galaxy likelihood function in Eq. (\ref{eq:like-single}) is the error-box defined by  $\Delta{\Omega}\times[z^{-}-\Delta{z^-_{v_p}}\, ,\, z^{+}+\Delta{z^+_{v_p}}]$. To perform the proposed statistical $D_l-z$ reconstruction, we now need to attach a realistic distribution of galaxy to each of those error-boxes. We use the simulated observed light cones of \cite{2012MNRAS.421.2904H} constructed from the Millennium run. The simulation assumes a $\Lambda$CDM cosmology with $h=0.73,\,\Omega_m=0.25,\,\Omega_\Lambda=0.75$ (which justifies the cosmological parameters assumed above to construct the BHB population). Because of the resolution of the simulation, galaxy distributions extracted from it are reliable only down to $M_*=10^{9}\msun$. We checked that at the magnitude cut imposed by \cite{2012MNRAS.421.2904H}, their reconstruction of the all-sky galaxy map is complete down to this mass limit out to $z\approx 0.15$. The all-sky galaxy map provides sky location, stellar mass, cosmological redshift $z$ and observed redshift $z_{\rm obs}$ (i.e. inclusive of peculiar velocities) for each galaxy. To select galaxy counterpart candidates to each GW event we proceed as follow:  
\begin{enumerate}
\item we fix the cosmological parameters to the values of the Millennium run and we select all galaxies with a {\it cosmological redshift} (i.e. excluding peculiar velocities) in the volume shell  $z\pm \Delta{z}$ consistent with the LISA measurement $D_l\pm \Delta{D_l}$;
\item we randomly choose one of those galaxies to be the 'true host' of the GW signal. Note that this amounts to say that the true host can be anywhere in the interval $D_l\pm \Delta{D_l}$, i.e., we are not centring the error-box around the true host;
\item we consider the error-box defined by  $\Delta{\Omega}\times[z^{-}-\Delta{z^-_{v_p}}\, ,\, z^{+}+\Delta{z^+_{v_p}}]$, where $\Delta{\Omega}$ is centred at the sky location of the selected 'true host';
\item we list all the galaxies with $M_*>10^{9}\msun$ that fall within the error-box. In doing this, we make sure that their {\it observed} redshift $z_{\rm obs}$ lies within the selected range. Each galaxy $z_{\rm obs}$ is the relevant $z_j$ that enters in Eq.~(\ref{eq:like-single}). In fact, in a real cosmological measurement one can only measure $z_{\rm obs}$ (which is affected by peculiar velocities) and does not have access to the true cosmological redshift.  
\end{enumerate}

\begin{figure}
\begin{center}
\includegraphics[width=7.5cm]{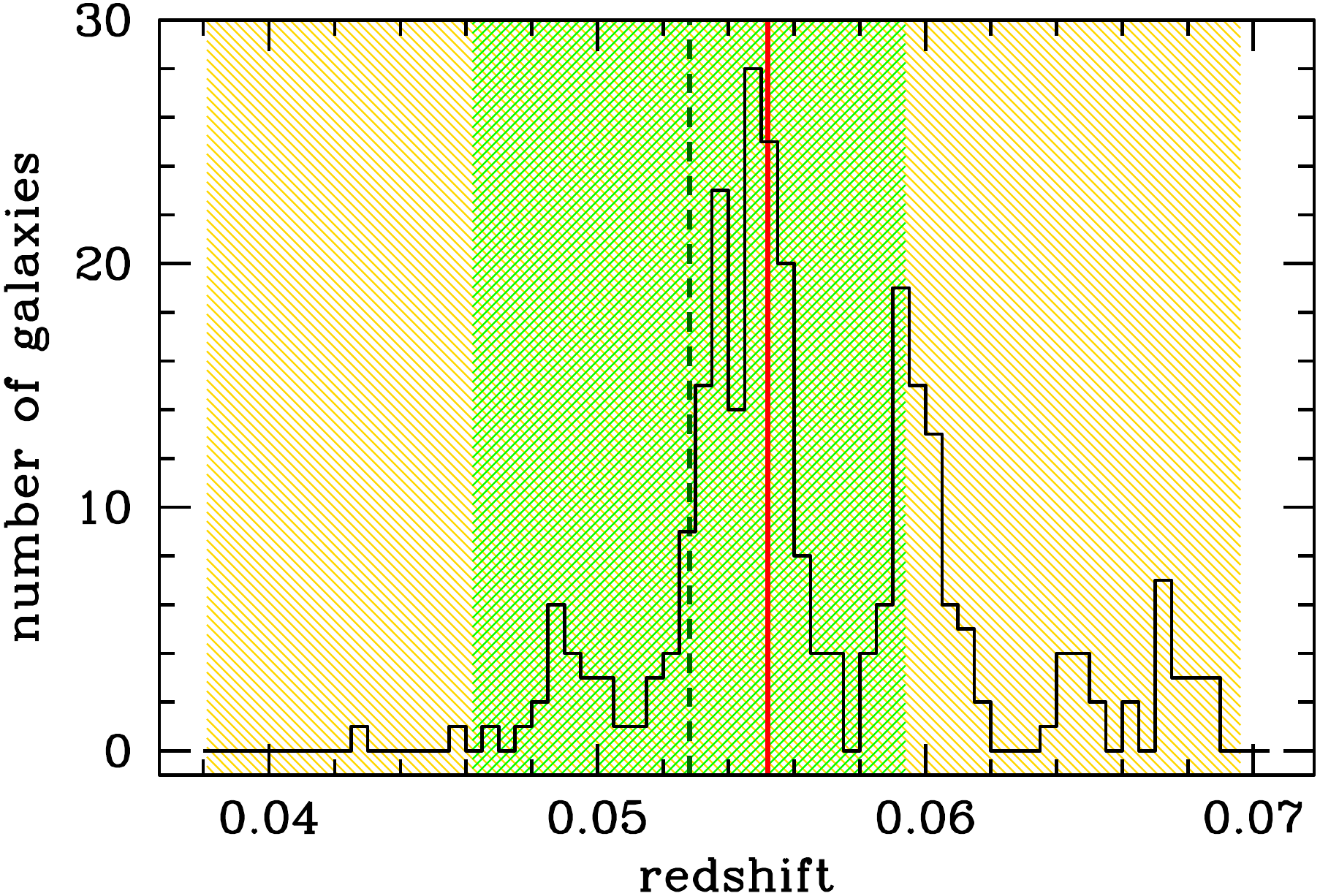}
\caption{Example of error-box construction and host selection. The red vertical line is the 'true host' redshift. The dashed vertical green line and the shaded green area are the best LISA estimated redshift and the 1$\sigma$ confidence interval, when the cosmology is fixed to the Millennium run values (i.e. to the true cosmology, in our experiment). The yellow area represents the full allowed redshift interval once prior uncertainty on the cosmological parameters and peculiar velocities are also taken into account, i.e. $[z^{-}-\Delta{z^-_{v_p}}\, ,\, z^{+}+\Delta{z^+_{v_p}}]$ as defined in the text. The black histogram is the redshift distribution of all possible galaxies falling in the error-box defined by $\Delta{\Omega}\times[z^{-}-\Delta{z^-_{v_p}}\, ,\, z^{+}+\Delta{z^+_{v_p}}]$.}
\label{fig:gal_dist}
\end{center}
\end{figure} 

The outcome of this procedure for a typical GW event is shown in Fig. \ref{fig:gal_dist}. Note that the procedure (and the statistical analysis described in Section \ref{sec:theo}) assigns to each galaxy equal probability of hosting the GW event. An alternative reasonable assumption would be that the probability of hosting the event is proportional to the stellar mass of the galaxy. This will likely sharpen the features of the likelihood function, leading to a better estimate of the cosmological parameters. In this respect, our procedure can be considered conservative and we defer the investigation of different assumptions to future work. 

In summary, for each LISA configuration and BHB population model, we construct a catalogue of $\vec{d}$ LISA events with $z<0.1$, observed with $S/N>8$ and a luminosity distance error $\Delta{D_l}/D_l<0.2$. Each BHB is assigned an array of possible galaxy hosts $j=1,...,m$ with ${z_{j,\rm obs}}$ consistent with the measured LISA uncertainty and the prior on the cosmological parameters. The distribution of number of possible hosts in the associated error-box defined by the GW measurement errors plus the prior in the cosmological parameters is shown in the bottom right panel of Fig. \ref{fig:pe}. Despite the small redshifts, the poor distance determination and the wide cosmology prior cause several galaxies to fall within the error-box $\Delta{\Omega}\times[z^{-}-\Delta{z^-_{v_p}}\, ,\, z^{+}+\Delta{z^+_{v_p}}]$. Differently from estimates by \cite{2016arXiv160907142K}, only few\% of the GW events (on average 9 for A5 and 2 for A2) have a single possible galaxy host, the vast majority have between 10 and several hundred possible hosts, making our statistical approach necessary.

Note also that we are considering hosts with $M_*>10^9\msun$ out to $z\approx 0.1$, and the number of galaxies within the error-box is generally less then a thousand. Contrary to the massive BHB case at cosmological distances investigated by \cite{2011ApJ...732...82P}, getting precise spectroscopic redshifts for all the host candidates will be a relatively straightforward task.

\section{Results and discussion}
\label{sec:results}
As shown in Fig.~\ref{fig:pe}, on average we expect about 20 and 130 useful BHBs in the A2 and A5 LISA baselines respectively. We investigate here three fiducial BHB population models corresponding to the $10-50-90$-th percentile of the rate distribution inferred from the first LIGO detections. The combination of detector baseline and population models results in 6 scenarios that we name A$2_{10}$, A$2_{50}$, A$2_{90}$, A$5_{10}$, A$5_{50}$, A$5_{90}$ and are listed in Tab. \ref{tab1}. For each scenario, we compute posterior distributions for $h$ and for the matter density $\Omega_m$. 
For each of the cases listed in Tab. ~\ref{tab1}, we generated 10 realisations of the set of possible galaxy hosts, thus all the results quoted in this section are averaged over the aforementioned realisations. 
Call $M$ the number of realisations for each event $d_i$, we produce average marginal likelihoods by computing
\begin{align}
p(<d_i>|h,\Omega_m,I) &= \frac{1}{M} \sum_{j=1}^{M} p(d_i^{j}|h,\Omega_m,I) =\nonumber\\
&= \frac{1}{M} \int dz \sum_{j=1}^{M} p(d_i^{j}|h,\Omega_m,z,I)
\end{align}
and $p(d_i^{j}|h,\Omega_m,z,I)$ for each event realisation is given in Eq.~\ref{eq:like-single}. The joint posteriors for $h$ and $\Omega_m$ are then computed from the averaged single ones. 
Regardless of the number of events considered (thus of the rate and instrumental configuration), we find that $\Omega_m$ is never measurable and its posterior is identical to the prior. 
This is not surprising as all the events we considered have a redshift smaller than 0.1. Hence, we will not report results for $\Omega_m$ as uninformative. 
\begin{table}
\begin{center}
\begin{tabular}{|ccccc|}
\hline
run & LISA design & ${\cal R}$ [yr$^{-1}$Gpc$^{-3}$] & $N_{\rm BHB}$ & $h$(68\%)\\
\hline
A$2_{10}$ & N2A2M5L6 & 12 &   7 & $0.716_{-0.050}^{+0.052}$\\
A$2_{50}$ & N2A2M5L6 & 34 &  22 & $0.734_{-0.033}^{+0.037}$\\
A$2_{90}$ & N2A2M5L6 & 70 &  39 & $0.726_{-0.024}^{+0.026}$\\
A$5_{10}$ & N2A5M5L6 & 12 &  55 & $0.730_{-0.020}^{+0.021}$\\
A$5_{50}$ & N2A5M5L6 & 34 & 143 & $0.728_{-0.012}^{+0.013}$\\
A$5_{90}$ & N2A5M5L6 & 70 & 259 & $0.731_{-0.009}^{+0.010}$\\
\hline
\hline 
\end{tabular}
\caption{For each of the instrumental configurations and coalescence rates considered, we report the number of sources observed as well as the 68\% credible intervals over $h$ averaged over 10 realisations of the galaxy hosts.}
\label{tab1}
\end{center}
\end{table}

\begin{figure}
\begin{center}
\includegraphics[width=7.5cm]{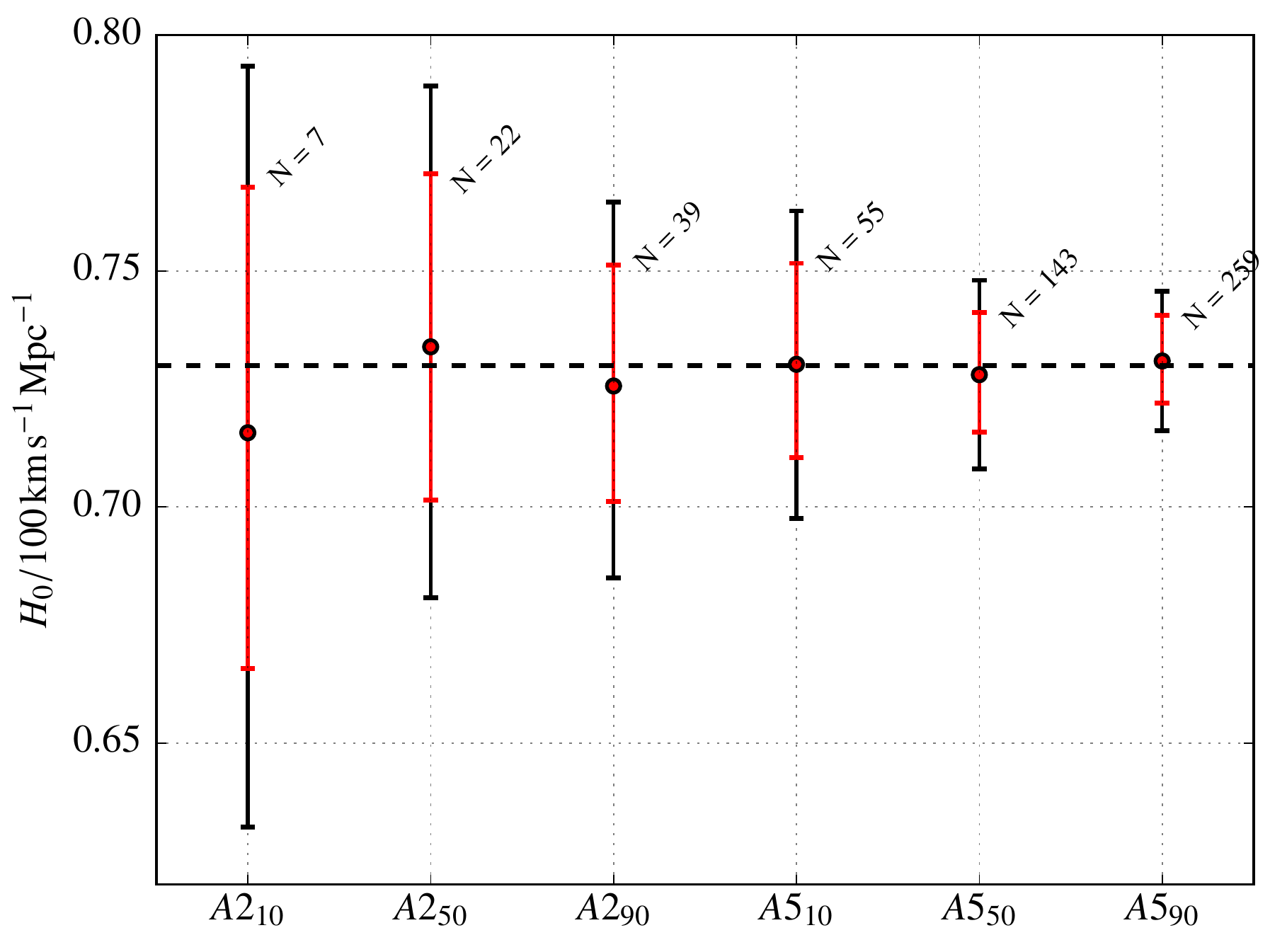}
\caption{90\% (black) and 68\% (red) credible intervals for $H_0/\mathrm{km}\,\mathrm{s}^{-1}\,\mathrm{Mpc}^{-1}$ for each of the configurations considered in this study. The credible regions are the result of the average over the galaxy hosts realisations. The main factor driving the inference of $H_0$ is the number of GW events detected, therefore the best inference is guaranteed by a high merger rate and 5Mkm armlength for LISA.}
\label{fig:summary}
\end{center}
\end{figure}

The main result of the study is summarised in Fig.~\ref{fig:summary}: LISA observations of BHBs will allow a determination of the Hubble constant $H_0$ with a degree of accuracy at the 68\% level that ranges from $\sim 7\%$ in the worst case scenario of a 2Mkm armlength and low merger rate to $\sim 1.3\%$ for the 5Mkm armlength and large merger rate. Rather than the increased S/N due to the longer baseline, we find that the inference is driven by the number of detected events. In fact, A$2_{90}$ and A$5_{10}$ yield approximately the same number of detections and consequently, very similar uncertainties on the Hubble constant, $3.5\%$ and $2.8\%$ respectively. The optimal inference for $h$ is given by the A$5_{90}$. In this case, the 259 detected events yield a measurement of $h$ whose 90\% accuracy is $\sim 2\%$, thus a factor of $\sim$2 worse than the current most accurate determination by the \citet{planck:2016}, but competitive with best distance indicators available today \citep{2016ApJ...826...56R}. While this is an unfair comparison, since LISA is still far in the future and the measurements provided either by the cosmic microwave background or by large SNe Ia monitoring campaigns are most likely going to improve substantially, we would like to stress that a GW based determination of the cosmological parameters is relying on entirely different assumptions and thus affected by independent systematics. This is in fact a key point; as mentioned in the introduction, some tension is emerging between CMB-based and local measurements of $H_0$. Before claiming hints of new physics, the possibility of systematics underestimation has to be ruled out. Local BHB standard sirens provide a direct benchmark comparison to SNe Ia measurements. A putative discrepancy between $H_0$ values measured with the two methods (SNe Ia and BHBs) will unveil an underestimation of the systematics at play in the measurements. On the contrary, a consistent measurement at odds with CMB-based inference, will strengthen the need of going beyond the standard $\Lambda$CDM model. 

Moreover, in this paper we are concerned exclusively with LISA observations of GW150914-like systems and not taking into account either the contributions coming from the analysis of LISA MBHBs \citep{2011ApJ...732...82P} and EMRIs \citep{2008PhRvD..77d3512M} or systems observed by LIGO/Virgo \citep{Nissanke2010a,2012PhRvD..86d3011D,Nissanke2013a} or Einstein Telescope \citep{Sathyaprakash2010a,2015arXiv150606590D}. Since the width of the posterior distribution for $h$ scales very close to $N_{\rm GW}^{-1/2}$, it is not unreasonable that by the end of the LISA mission the accuracy with which $h$ can be measured using GW observations alone will be competitive with the measurements coming from more traditional channels. 

\begin{figure*}
\begin{center}
\includegraphics[width=7.5cm]{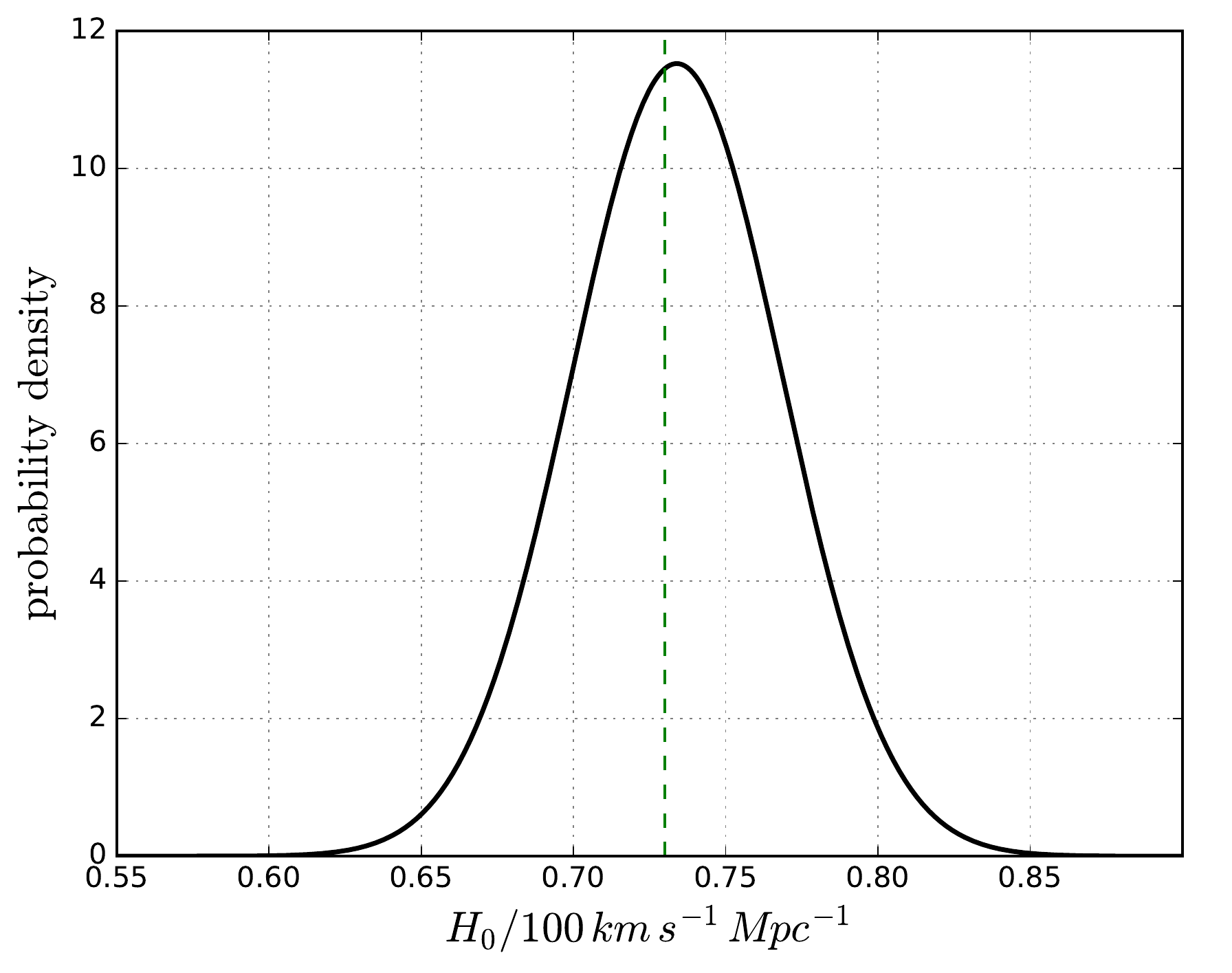} 
\includegraphics[width=7.5cm]{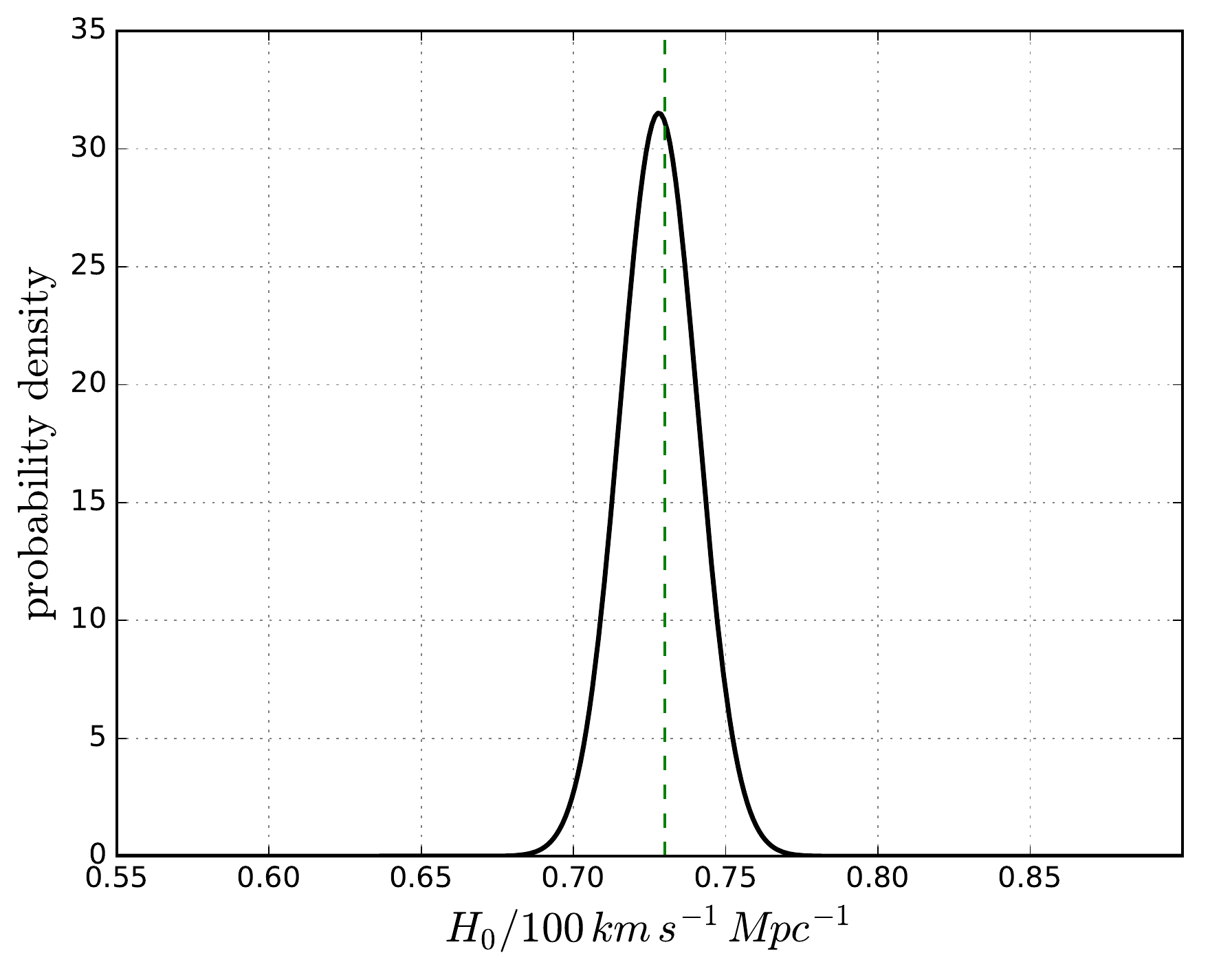} 
\includegraphics[width=7.5cm]{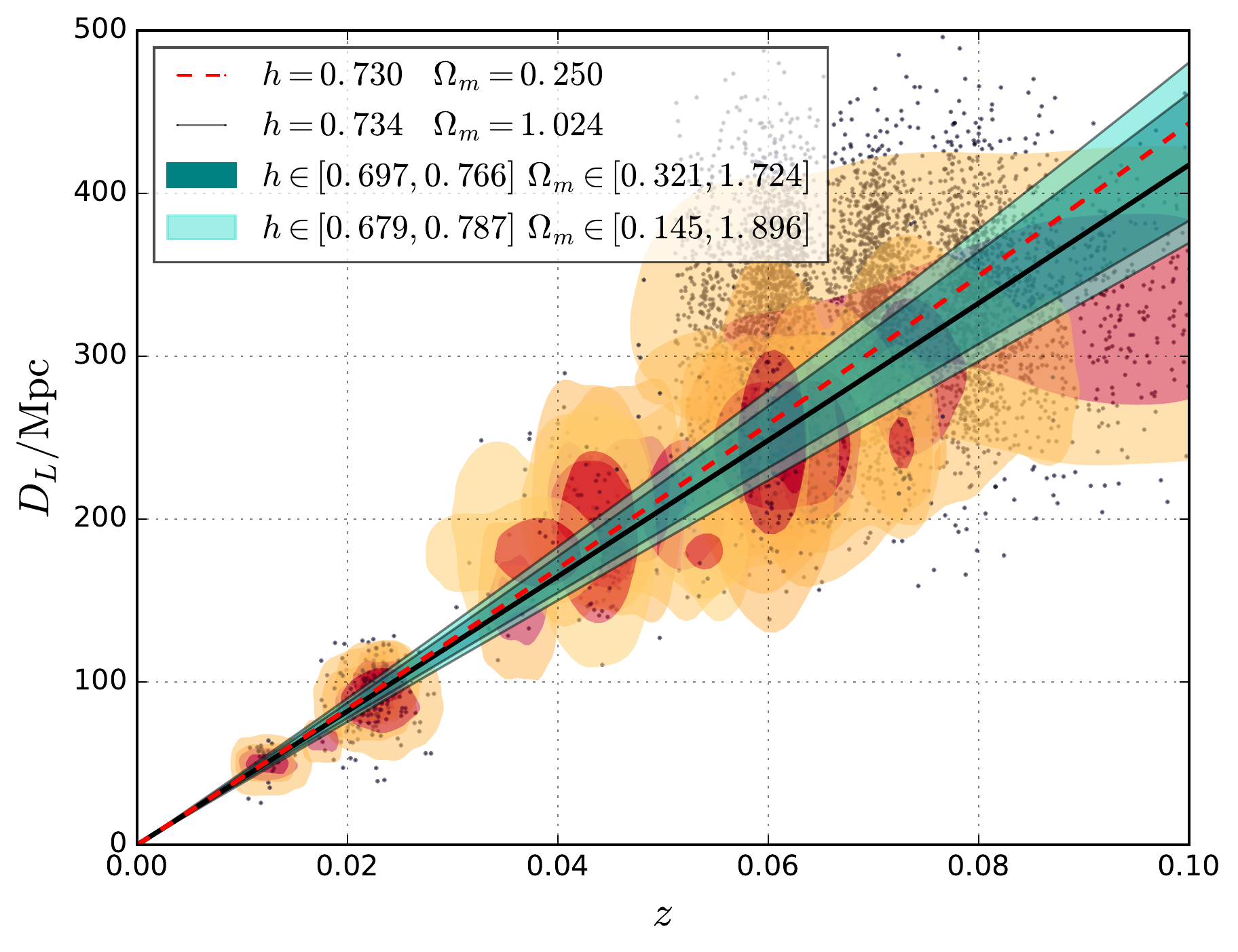}
\includegraphics[width=7.5cm]{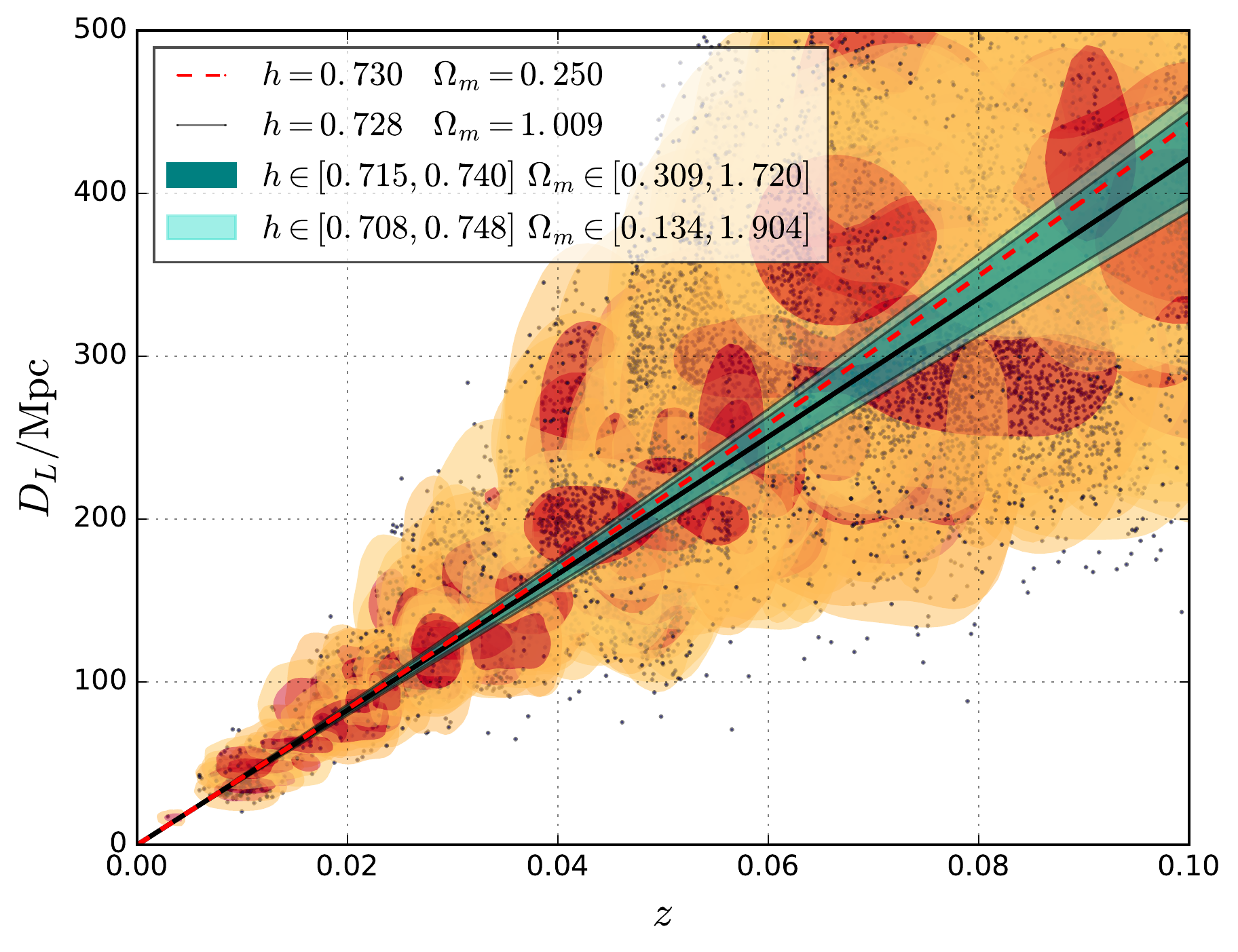}
\includegraphics[width=7.5cm]{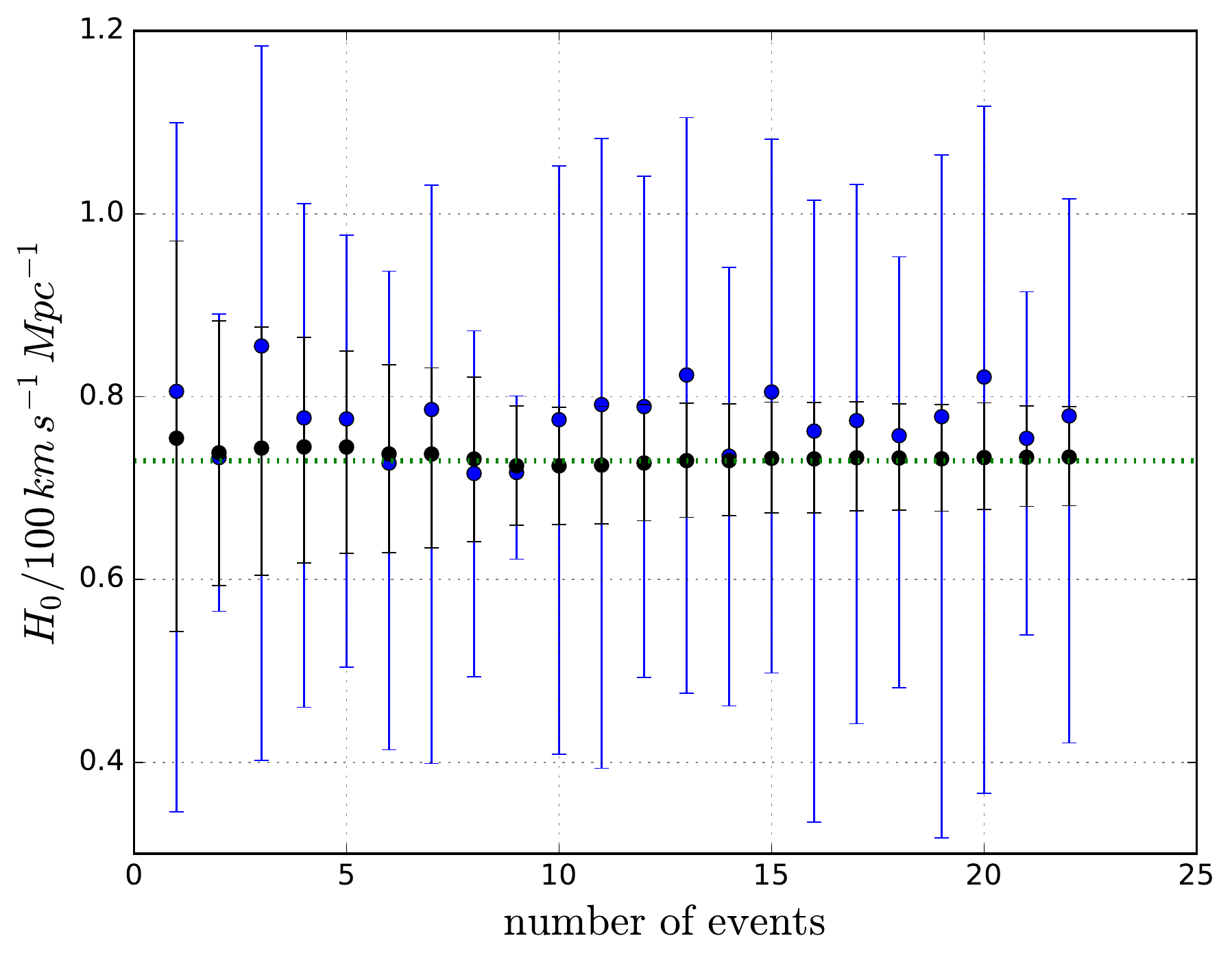}
\includegraphics[width=7.5cm]{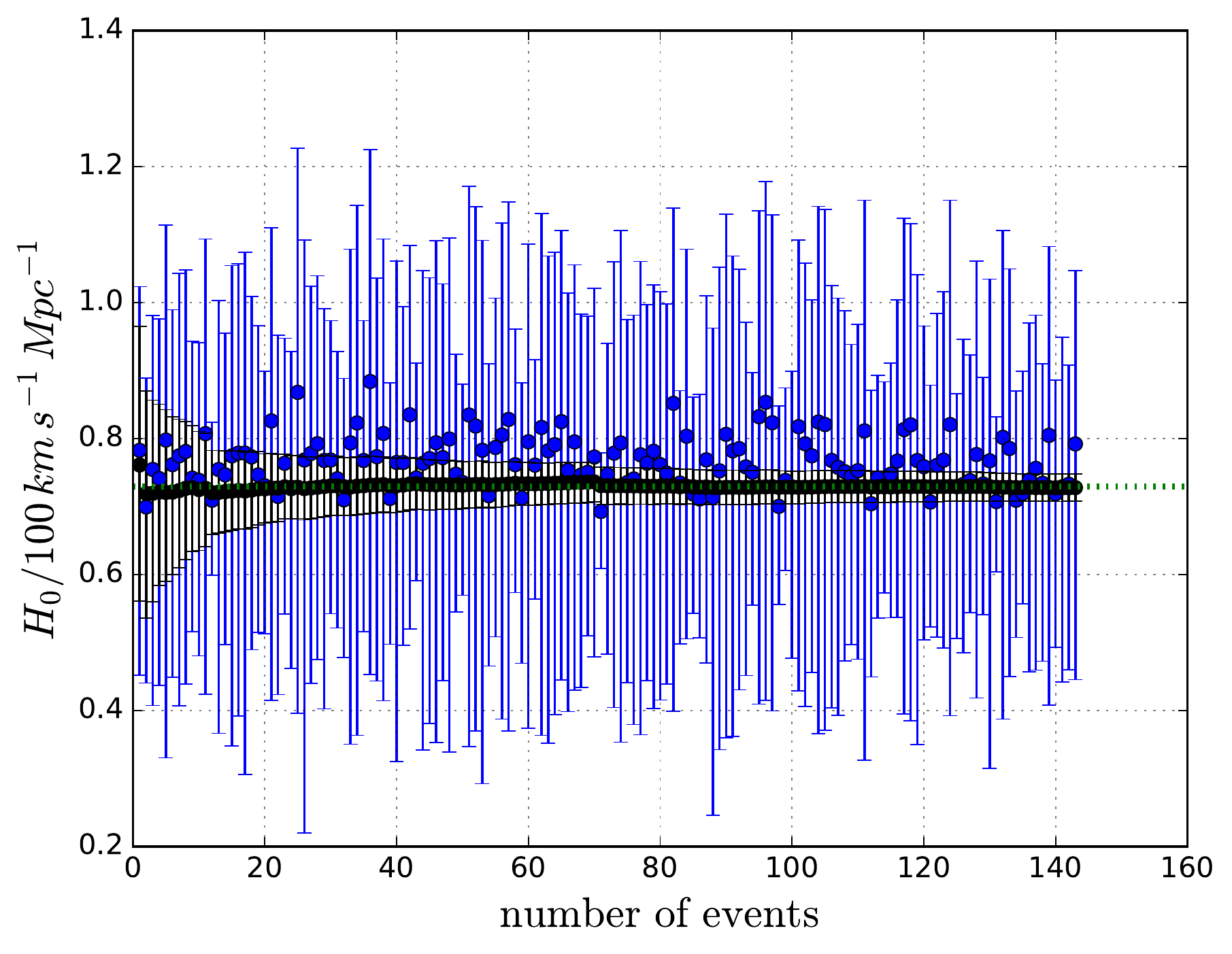}
\caption{Example of the inference for A$2_{50}$ (left column) and A$5_{50}$ (right column). \emph{Top row}: joint posterior distributions for $h$. The green dashed lines indicate the fiducial value $h=0.73$. \emph{Centre row}: median $D_L - z$ regression line (solid) and $68\%$ and $90\%$ credible regions in teal and turquoise, respectively. Each set of contours indicate the 50th, 90th percentiles from the single event posterior distribution for the redshift $z$. The dashed line indicates the  $D_L - z$  relation from the fiducial values of the cosmological panels, $h=0.73$ and $\Omega_m=0.25$. The black dots are the redshifts and luminosity distances -- computed using the fiducial cosmology -- for the individual galaxies that went in the analysis. \emph{Bottom row}: evolution of the $90\%$ credible region for $h$ as a function of the number of events analysed (black error bars). In blue, the $90\%$ credible region for $h$ from the analysis is the corresponding individual event. The green dotted line indicates the fiducial value $h=73$. }
\label{fig:example-pdf}
\end{center}
\end{figure*} 

Our results are moderately affected by the specific realisation of the galactic hosts. For each of the cases here considered, we find that the variance of the median value of $h$ inferred is similar to the average joint posterior variance. The width of the joint posterior is, not unexpectedly, unaffected. The latter depends, in fact, only on the sizes of the distance uncertainties and not on the details of the galactic population. Examples of individual realizations of the observed standard sirens sample and on the associated $H_0$ constrains are shown in Fig.~\ref{fig:example-pdf} for the A$2_{50}$ and A$5_{50}$ cases. The top panels show that the width of the joint posteriors of the single realizations is in fact very similar to the average values reported in Tab. ~\ref{tab1}, whereas the bottom panels highlight the contribution of each individual BHB to the $H_0$ measurement and the $N_{\rm GW}^{-1/2}$ improvement in the joint $H_0$ constrain as systems are added up. The central panels provide a visual representation of how our technique works. Each BHB has an associated 'likelihood blob' in the $D_L-z$ plane given by the uncertainty in the GW measurement, weighted with the distribution of galaxies within this uncertainty. As we add up blobs, we can find the cosmological parameters that provide the best $D_L-z$ fit to them. Note that systems at $z<0.05$ have significantly smaller associated uncertainties, contributing the most to the measurement. However, at such small distances, galaxy peculiar velocities can significantly bias the estimate of the intrinsic redshift of the host galaxies, and it is important to average over several events to avoid a significant bias in the final measurement of $H_0$. Moreover, similarly to what observed for EM measurements from the  local Universe \citep[e.g][]{2016ApJ...826...56R}, the actual measurement of $H_0$ from the LISA sources using the method presented in this study will have to assess potential systematics caused by cosmic variance effects which can have a non-negligible impact (of the order of 1\%) over the inferred value of $H_0$. 

Finally, from the class of sources considered here, it is not possible to draw any inference over the value of the matter density $\Omega_m$. This is because the redshift range we considered is limited to $z\leq 0.1$, over which the $D_L - z$ relation is very little affected. For an inference of the energy density parameters with LISA observations, we would have to turn to massive BH binaries which can be detected at far larger redshifts and indeed have the potential of probing even the dark energy equation of state \citep[e.g.][]{2011ApJ...732...82P}. The lack of sensitivity to $\Omega_m$, however, greatly simplified our analysis. For searches that are sensitive to $\Omega_m$ and whose indicators have large distance uncertainties, selection effects can play a significant role \citep[see discussion in ][]{2015arXiv150606590D}. Therefore, a simultaneous inference of the coalescence rate (irrelevant for the case of $H_0$) and of the cosmological parameters is necessary. 

\section{Conclusions}
\label{sec:conclusions}

We considered the possibility of using GW150914-like sources as observed by LISA as cosmological probes. Building on the original idea of \cite{2008PhRvD..77d3512M}, we emploied a statistical methodology that circumvent the need of direct EM counterparts to the GW events to populate the $D_l-z$ diagram. Essentially, the method associates to each BHB a redshift probability distribution defined by the galaxies falling in an error cube given by the GW measurement uncertainty plus a prior uncertainty range for the cosmological parameters. A measurement of the latter is then possible by stacking several such distributions  and searching for the combination of cosmological parameters that maximizes the resulting joint probability of the BHB hosts.

We demonstrated the feasibility of such measurement by performing end-to-end simulations including: i. state of the art rates of BHB mergers as inferred by the first Advanced LIGO run; ii. sky localization and distance determination of each BHB as determined by LISA observations employing two different baselines and the full response function of the detector; iii. realistic distributions of potential galaxy hosts taken from the light cones derived from the Millennium simulation. 

We found that LISA can probe $H_0$ with an accuracy that ranges from $\sim$7\% to $\sim$1\% (68\% confidence) depending on the intrinsic BHB merger rate and detector configuration. Except for the most pessimistic cases, figures are comparable with current local measurement of $H_0$ and can therefore play an important role in addressing the apparent discrepancy between local and CMB based measurements of the Hubble constant. Although EM-based $H_0$ measurements are expected to improve by the time LISA flies, we stress that GW based methods rely on independent assumptions and systematics. Moreover, LISA measurement of $H_0$ from BHB can be combined with results from other classes of GW standard sirens (massive BHBs, EMRIs, etc.) to significantly reduce the associated uncertainties.

While conducting this study, a novel LISA baseline has been proposed by the LISA Consortium to address the L3 mission call by ESA \citep{2017arXiv170200786A}. The new baseline features a 2.5 Mkm armlength and slightly different technical specifications. The resulting sensitivity curve is similar to the N2A2M5L6 configuration investigated here. The minimum proposed mission lifetime is four years, with the goal of an extension to up to ten years. \cite{2017arXiv170204356S} derived BHB observed rates for the new configuration, confirming that they are in line with the N2A2M5L6 baseline. If the mission is extended to ten years, however, the number of BHB detections increases by a factor of $\approx 4$, implying a factor of $\approx 2$ better determination of $H_0$. Since the determination of $H_0$ in the N2A2M5L6 case ranges from $7\%$ to $3.5\%$, a ten year mission with the new proposed baseline can push these figures down to $3\%$ to $1.5\%$, depending on the intrinsic BHB merger rate.

Finally, we caution that, although quite sophisticated, our simulations rely on a number of simplifying assumptions. First, GW measurement errors are computed in the FIM approximation. Although this might not be accurate for relatively low S/N events, results from the mock LISA data challenge \citep{2010CQGra..27h4009B} showed that FIM results are quite reliable also at low S/N for persistent sources like EMRIs or galactic WD binaries. Second, we only consider the $1\sigma$ GW measurement error-cube and assign to each galaxy within the cube equal probability to host the BHB. In reality, one will have to weight the probability of each galaxy according to the shape of the 3D joint posterior distribution of the source distance and sky location. Lastly, our host probability distribution might be refined by applying more physically motivated weights for each galaxy. For example, the probability of hosting a BHB can be taken to be proportional to the galaxy mass. This would likely produce more defined spikes in the host probability distributions, with a consequential improvement on the determination of the cosmological parameters.


\section*{Acknowledgements}
The simulations were performed on the Tsunami cluster of the University of Birmingham.
A.S. is supported by a University Research Fellowship of the Royal Society. WDP is supported by the program ``Rientro dei Cervelli Rita Levi Montalcini".
A. K. is supported by NSF CAREER Grant No. PHY-1055103,
and by FCT contract IF/00797/2014/CP1214/CT0012
under the IF2014 Programme. This work was supported
by the Centre National d?{\'E}tudes Spatiales.
Part of the graphs on this manuscript have been produced using \texttt{Matplotlib}~\citep{Hunter:2007}.

\bibliographystyle{mnras}
\bibliography{bibnote}

\bsp

\label{lastpage}

\end{document}